\documentclass[superscriptaddress,twocolumn,showpacs,aps,pre]{revtex4-1}
\usepackage{amssymb,amsmath}
\usepackage[dvips]{graphicx}

\begin{document}

\title{Strong-randomness phenomena in quantum Ashkin-Teller models}

\author{Hatem Barghathi}
\affiliation{Department of Physics, Missouri University of Science and Technology, Rolla, MO 65409, USA}

\author{Fawaz Hrahsheh}
\affiliation{Department of Physics, Missouri University of Science and Technology, Rolla, MO 65409, USA}
\affiliation{Department of Physics, Jordan University of Science and Technology, Irbid 22110, Jordan}
\affiliation{King Fahd University of Petroleum and Minerals, Dhahran 31261, Saudi Arabia}

\author{Jos\'e A. Hoyos}
\affiliation{Instituto de F\'{i}sica de S\~ao Carlos, Universidade de S\~ao Paulo,
C.P. 369, S\~ao Carlos, S\~ao Paulo 13560-970, Brazil}

\author{Rajesh Narayanan}
\affiliation{Department of Physics, Indian Institute of Technology Madras, Chennai 600036, India}

\author{Thomas Vojta}
\affiliation{Department of Physics, Missouri University of Science and Technology, Rolla, MO 65409, USA}

\begin{abstract}
The $N$-color quantum Ashkin-Teller spin chain is a prototypical model for the study of
strong-randomness phenomena at first-order and continuous quantum phase transitions.
In this paper, we first review the existing strong-disorder renormalization group approaches to
the random quantum Ashkin-Teller chain in the weak-coupling as well as the strong-coupling
regimes.
We then introduce a novel general variable transformation that unifies the treatment of the strong-coupling
regime. This allows us to determine the phase diagram for all color numbers $N$, and the critical
behavior for all $N \ne 4$.
In the case of two colors, $N=2$, a partially ordered product phase separates the paramagnetic and ferromagnetic
phases in the strong-coupling regime.
This phase is absent for all $N>2$, i.e., there is a direct phase boundary between the paramagnetic and ferromagnetic
phases.
In agreement with the quantum version of the Aizenman-Wehr theorem, all phase transitions are continuous,
even if their clean counterparts are of first order.  We also discuss the various critical and multicritical
points. They are all of infinite-randomness type, but depending on the coupling strength, they
belong to different universality classes.
\end{abstract}

\date{\today}

\maketitle

\section{Introduction}
\label{sec:intro}

Simple models of statistical thermodynamics have played a central role in our understanding
of phase transitions and critical phenomena. For example, Onsager's solution of the two-dimensional
Ising model \cite{Onsager44} paved the way for the use of statistical mechanics methods
in the physics of thermal (classical) phase transitions. More recently, the transverse-field Ising chain
has played a similar role for quantum phase transitions \cite{Sachdev_book99}.

The investigation of systems with more complex phase diagrams requires richer models. For example, the
quantum Ashkin-Teller spin chain \cite{AshkinTeller43,KohmotoNijsKadanoff81,IgloiSolyom84}
and its $N$-color generalization \cite{GrestWidom81,Fradkin84,Shankar85} feature partially ordered
intermediate phases, various first-order and continuous quantum phase transitions, as well as
lines of critical points with continuously varying critical exponents.
Recently, the quantum Ashkin-Teller
model has reattracted considerable attention because it can serve as a prototypical model
for the study of various strong-randomness effects predicted to occur at quantum phase transitions
in disordered systems \cite{Vojta06,Vojta10}.

In the case of $N=2$ colors, the correlation length exponent $\nu$ of the clean quantum Ashkin-Teller
model varies
continuously with the strength of the coupling between the colors. The disorder can therefore be tuned from being
perturbatively irrelevant (if the Harris criterion \cite{Harris74} $d\nu>2$ is fulfilled) to relevant
(if the Harris criterion is violated). For more than two colors, the clean system features
a first-order quantum phase transition. It is thus a prime example for exploring the effects
of randomness on first-order quantum phase transitions and for testing the predictions of the
(quantum) Aizenman-Wehr theorem \cite{AizenmanWehr89,GreenblattAizenmanLebowitz09}.

In this paper, we first review the physics of the random quantum Ashkin-Teller chain
in both the weak-coupling and the strong-coupling regimes, as obtained by various
implementations of the strong-disorder renormalization group.
We then introduce a variable transformation scheme that permits a unified treatment of the
strong-coupling regime for all color numbers $N$. The paper is organized as follows:
The Hamiltonian of the $N$-color quantum Ashkin-Teller chain is introduced in Sec.\
\ref{sec:AT}. Section \ref{sec:weak_coupling} is devoted to disorder phenomena in the
weak-coupling regime.  To address the strong-coupling regime in Sec.\ \ref{sec:strong_coupling},
we first review the existing results and then  introduce a general variable transformation. We also discuss the resulting phase diagrams
and phase transitions. We conclude in Sec.\ \ref{sec:conclusions}.

\section{$N$-Color quantum Ashkin-Teller chain}
\label{sec:AT}

The one-dimensional $N$-color quantum Ashkin-Teller model  \cite{GrestWidom81,Fradkin84,Shankar85} consists of $N$ identical
transverse-field Ising chains of length $L$ (labeled by the ``color'' index $\alpha=1\ldots N$) that
are coupled via their energy densities. It is given by the Hamiltonian
\begin{eqnarray}
\label{eq:HAT}
 H=&-&\sum_{\alpha=1}^N\sum_{i=1}^L{\left ( J_i S_{\alpha,i}^z S_{\alpha,i+1}^z + h_i S_{\alpha,i}^x \right )} \\
&-&\sum_{\alpha<\beta}\sum_{i=1}^L{\left (K_i S_{\alpha,i}^z S_{\alpha,i+1}^z S_{\beta,i}^z S_{\beta,i+1}^z + g_i S_{\alpha,i}^x S_{\beta,i}^x\right )} ~. \nonumber
\end{eqnarray}
$S_{\alpha,i}^x$  and $S_{\alpha,i}^z$ are Pauli matrices that describe the spin of color $\alpha$ at lattice site $i$.
The strength of the inter-color coupling can be characterized by the ratios $\epsilon_{h,i}=g_i/h_i$ and $\epsilon_{J,i}=K_i/J_i$.
In addition to its fundamental interest, the Ashkin-Teller model has been applied to
absorbed atoms on surfaces \cite{Baketal85}, organic magnets, current loops in
high-temperature superconductors \cite{AjiVarma07,AjiVarma09} as well as the elastic response of DNA
molecules  \cite{ChangWangZheng08}.
The quantum Ashkin-Teller chain (\ref{eq:HAT}) is invariant under the duality transformation
$S_{\alpha,i}^z S_{\alpha,i+1}^z \to \tilde S_{\alpha,i}^x$,
$S_{\alpha,i}^x \to \tilde S_{\alpha,i}^z \tilde S_{\alpha,i+1}^z$,
 $J_i\rightleftarrows h_i$, and $\epsilon_{J,i}\rightleftarrows\epsilon_{h,i}$,
where $\tilde S_{\alpha,i}^x$ and $\tilde S_{\alpha,i}^z$ are the dual Pauli matrices
\cite{Baxter_book82}.
This self-duality symmetry will prove very useful in fixing the positions of various
phase boundaries of the model.

In the clean problem, the interaction energies and fields are uniform in space,
$J_i\equiv J$, $K_i \equiv K$, $h_i\equiv h$, $g_i\equiv g$, and so are the coupling
strengths $\epsilon_{h,i}\equiv \epsilon_{h}$ and $\epsilon_{J,i}\equiv \epsilon_{J}$.
In the present paper, we will be interested in the effects of quenched disorder.
We therefore take the interactions $J_i$ and transverse fields $h_i$ as independent
random variables with probability distributions $P_0(J)$ and $R_0(h)$.
$J_i$ and $h_i$ can be restricted to positive values, as possible negative signs can
be transformed away by a local transformation of the spin variables. Moreover, we focus on the
case of nonnegative couplings, $\epsilon_{J,i}, \epsilon_{h,i} \ge 0$. In most of the paper
we also assume that the coupling strengths in the bare Hamiltonian (\ref{eq:HAT}) are spatially uniform,
$\epsilon_{J,i}=\epsilon_{h,i} = \epsilon_I$.
Effects of random coupling strengths will be considered in the concluding section.

\section{Weak coupling regime}
\label{sec:weak_coupling}

For weak coupling and weak disorder, one can map the Ashkin-Teller model onto a continuum
field theory and study it via a perturbative renormalization group
\cite{Dotsenko85,GiamarchiSchulz88,GoswamiSchwabChakravarty08}. This renormalization
group displays runaway-flow towards large disorder indicating a breakdown of the perturbative
approach. Consequently, nonperturbative methods
are required even for weak coupling.

Carlon et al.\ \cite{CarlonLajkoIgloi01} therefore investigated the weak-coupling
regime $|\epsilon_I|<1$ of the two-color random quantum Ashkin-Teller chain
using a generalization of Fisher's strong-disorder renormalization group \cite{Fisher92,Fisher95}
of the random transverse-field Ising chain.
Analogously, Goswami et al.\ \cite{GoswamiSchwabChakravarty08}
considered the $N$-color version for $0\le \epsilon_I < \epsilon_c(N)$ where
$\epsilon_c$ is an $N$-dependent constant. In the following, we summarize their
results to the extent necessary for our purposes, focusing on nonnegative $\epsilon_I$.

The bulk phases of the random quantum Ashkin-Teller model (\ref{eq:HAT})
in the weak-coupling regime are easily understood.
If the interactions $J_{i}$ dominate over the fields $h_{i}$,
the system is in the ordered (Baxter) phase in which each color orders ferromagnetically.
In the opposite limit, the model is in the paramagnetic phase.

The idea of any strong-disorder renormalization group method consists in finding the
largest local energy scale and integrating out the corresponding high-energy degrees of
freedom. In the weak-coupling random quantum Ashkin-Teller model, the largest local energy
is either a transverse field $h_i$ or an interaction $J_i$. We thus set the high-energy
cutoff of the renormalization group to $\Omega=\max(h_i,J_i)$. If the largest energy is
the transverse-field $h_i$, the local ground state $|\to\to \ldots \to\rangle$ has all spins
at site $i$ pointing in the positive $x$ direction (each arrow represents one color).
Site $i$ thus does not contribute to the order parameter, the $z$-magnetization, and can be
integrated out in a site decimation step. This leads to effective interactions between sites $i-1$ and $i+1$.
Specifically, one obtains an effective Ising interaction
\begin{equation}
\tilde J = \frac {J_{i-1} J_{i}}{h_i +(N-1) g_i}~
\label{eq:SDRG_Jeff}
\end{equation}
and an effective four-spin interaction
\begin{equation}
\tilde K = \frac {K_{i-1} K_{i}}{2 [h_i + (N-2) g_i]}~.
\label{eq:SDRG_Keff}
\end{equation}
This implies that the coupling strength $\epsilon$ renormalizes as
\begin{equation}
\tilde \epsilon_J = \frac {\epsilon_{J,i-1}\epsilon_{J,i}} 2 ~\frac{1+(N-1)\epsilon_{h,i}}{1+(N-2)\epsilon_{h,i}}~.
\label{eq:SDRG_eps_J}
\end{equation}
The recursion relations for the case of the largest local energy being the interaction $J_i$
can be derived analogously or simply inferred from the self-duality of the Hamiltonian. In this case,
the sites $i$ and $i+1$ are merged into a single new site whose fields and coupling strength are given by
\begin{equation}
\tilde h = \frac {h_{i} h_{i+1}}{J_i +(N-1) K_i}~,
\label{eq:SDRG_heff}
\end{equation}
\begin{equation}
\tilde g = \frac {g_{i} g_{i+1}}{2 [J_i + (N-2) K_i]}~,
\label{eq:SDRG_geff}
\end{equation}
\begin{equation}
\tilde \epsilon_h = \frac {\epsilon_{h,i}\epsilon_{h,i+1}} 2 ~\frac{1+(N-1)\epsilon_{J,i}}{1+(N-2)\epsilon_{J,i}}~.
\label{eq:SDRG_eps_h}
\end{equation}
According to eqs.\ (\ref{eq:SDRG_eps_J}) and (\ref{eq:SDRG_eps_h}), the coupling strengths $\epsilon$ renormalize downward
without limit under the strong-disorder renormalization group provided their initial values are sufficiently small. Assuming a uniform
initial $\epsilon_I$, the coupling strength decreases if $\epsilon_I < \epsilon_c(N)$ with the critical value
given by
\begin{equation}
\epsilon_c(N) = \frac {2N-5}{2N-2} + \sqrt{\left(\frac {2N-5}{2N-2}\right)^2 + \frac{2}{N-1}}~.
\label{eq:epsilon_c}
\end{equation}
It takes the value
$\epsilon_c(2)=1$ and increases monotonically with $N$ towards the limit $\epsilon_c(\infty)=2$.

If $\epsilon_I < \epsilon_c(N)$, the $N$ random transverse-field Ising chains making up the random quantum Ashkin-Teller model
asymptotically decouple. The low-energy physics of the random quantum Ashkin-Teller model is thus identical to that of the
random transverse-field Ising chain. In particular, there is a direct quantum phase transition between the ferromagnetic and paramagnetic
phases. In agreement with the self-duality of the Hamiltonian, it is located at $J_{\rm typ} = h_{\rm typ}$ where the typical values
$J_{\rm typ}$ and  $h_{\rm typ}$ are be defined as the geometric means of the distributions $P_0(J)$ and $R_0(h)$.
The critical behavior of the transition is of infinite-randomness
type and in the random transverse-field Ising universality class  \cite{Fisher95}. It is accompanied by power-law quantum Griffiths singularities.

\section{Strong coupling regime}
\label{sec:strong_coupling}
\subsection{Existing results}

If  $\epsilon_I > \epsilon_c(N)$, the coupling strengths increase under the renormalization group steps of Sec.\
\ref{sec:weak_coupling}. If they get sufficiently large, the energy spectrum of the local Hamiltonian changes, and the
method breaks down. To overcome this problem, two recent papers have implemented versions of the strong-disorder
renormalization group that work in the strong-coupling limit $\epsilon \to \infty$
\cite{HrahshehHoyosVojta12,HHNV14}.

For large $\epsilon$, the inter-color couplings
in the second line of the Hamiltonian (\ref{eq:HAT}) dominate over the Ising terms in the first line. The
low-energy spectrum of the local Hamiltonian therefore consists of a ground-state sector and a
pseudo ground-state sector, depending on whether or not a state satisfies the Ising terms \cite{HrahshehHoyosVojta12}.
For different numbers of colors $N$, this leads to different consequences.

For $N>4$, the local binary degrees of freedom that distinguish the two sectors become asymptotically free in the low-energy limit.
By incorporating them into the strong-disorder renormalization group approach, the authors of Ref.\
\cite{HrahshehHoyosVojta12} found that the direct continuous quantum transition between the ferromagnetic and paramagnetic phases
on the self-duality line $J_{\rm typ} = h_{\rm typ}$ persists in the strong-coupling regime
$\epsilon_I > \epsilon_c(N)$. In agreement with the quantum Aizenman-Wehr theorem \cite{GreenblattAizenmanLebowitz09}, the first order transition of the clean
model is thus replaced  by a continuous one.  However, the critical behavior in the strong-coupling regime differs from
the random transverse-field Ising universality class that governs the weak-coupling case. The critical point is still of
infinite-randomness type, but the additional degrees of freedom lead to even stronger thermodynamic singularities.
The method of Ref.\  \cite{HrahshehHoyosVojta12} relies on the ground-state and pseudo ground-state sectors decoupling
at low energies and thus holds for $N>4$ colors only.

\begin{figure}[t]
\includegraphics[width=8.5cm]{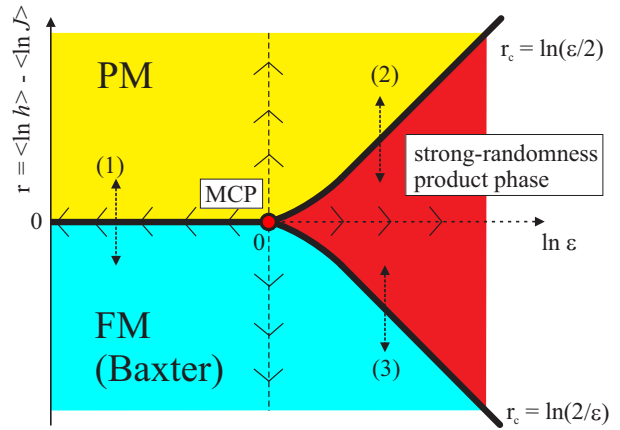}
\caption{Schematic ground state phase diagram of the two-color random quantum Ashkin-Teller chain. For $\epsilon_I<\epsilon_c(1)=1$, the
paramagnetic and ferromagnetic phases are connected by a direct continuous quantum phase transition. For
$\epsilon_I>1$, they are separated by a partially ordered ``product'' phase characterized by strong randomness and renormalization
group flow towards infinite coupling. The two regimes are separated by a multicritical point (MCP) at
$\epsilon=1$. (after Ref.\ \cite{HHNV14}).}
\label{fig:2AT_PD}
\end{figure}

We now turn to $N=2$.
The strong-coupling regime of the two-color random quantum Ashkin-Teller model was recently attacked \cite{HHNV14} by the
variable transformation
\begin{eqnarray}
\sigma_i^z &=& S_{1,i}^z S_{2,i}^z~, \quad \eta_i^z = S_{1,i}^z~
\label{eq:2AT_trans_z}
\end{eqnarray}
which introduces the product of the two colors as an independent variable.
The corresponding transformations for the Pauli matrices
$S_{1,i}^x$ and $S_{2,i}^x$ read
\begin{eqnarray}
\sigma_i^x &=& S_{2,i}^x~, \quad \eta_i^x = S_{1,i}^x S_{2,i}^x~.
\label{eq:2AT_trans_x}
\end{eqnarray}
Inserting these transformations into the  $N=2$ version of the Hamiltonian (\ref{eq:HAT}) gives
\begin{eqnarray}
H = &-& \sum_i (K_i \sigma_i^z\sigma_{i+1}^z + h_i \sigma_i^x) - \sum_i (J_i \eta_i^z\eta_{i+1}^z + g_i \eta_i^x)  \nonumber\\
    &-& \sum_i (J_i\sigma_i^z\sigma_{i+1}^z\eta_i^z\eta_{i+1}^z + h_i \sigma_i^x\eta_i^x)~.
\label{eq:Hprod}
\end{eqnarray}
An intuitive physical picture of the strong-coupling regime $\epsilon \gg 1$ close to
self duality, $h_{\rm typ} \approx J_{\rm typ}$, emerges directly from this Hamiltonian.
The  product variable $\sigma$ is dominated by the four-spin interactions $K_i$ while the behavior of
the variable $\eta_i$ which traces the original spins is dominated by the two-spin transverse fields $g_i$.
All other terms vanish in the limit $\epsilon \to \infty$, i.e., the pair product variable and the spin variable
asymptotically decouple. The system is therefore in a partially ordered phase
in which the pair product variable $\sigma^z$ develops long-range order while the spins remain disordered.
A detailed strong-disorder renormalization group study \cite{HHNV14} confirms this picture
and also yields the complete phase diagram (see Fig. \ref{fig:2AT_PD}) as well as the critical behaviors of the
various quantum phase transitions.
For example, the transitions between the product phase and the paramagnetic and ferromagnetic phases (transitions 2 and 3 in Fig.\ \ref{fig:2AT_PD})
are both of infinite-randomness type and in the random transverse-field Ising universality class.

The strong-coupling behavior of the random quantum Ashkin-Teller chains with $N=3$ and 4 colors could not be worked out with the above methods.

\subsection{Variable transformation for $N=3$}

In this and the following subsections, we present a method that allows us to study the strong-coupling regime of the random quantum
Ashkin-Teller model for any number $N$ of colors. It is based on a generalization of the variable transformation
(\ref{eq:2AT_trans_z}), (\ref{eq:2AT_trans_x}) of the two-color problem.
We start by discussing $N=3$ colors which is particularly interesting because it is not covered by the existing
work \cite{HrahshehHoyosVojta12,HHNV14}. Furthermore, it is the lowest number of colors for which  the clean system
features a first-order transition. After $N=3$, we consider general odd and even color numbers $N$ which require
slightly different implementations.

In the three-color case, the transformation is defined by introducing two pair variables and one product of all three original colors,
\begin{equation}
\sigma_{i}^z = S_{1,i}^z\,S_{3,i}^z ,\quad \tau_{i}^z = S_{2,i}^z\,S_{3,i}^z, \quad
\eta_{i}^z =  S_{1,i}^z\,S_{2,i}^z\,S_{3,i}^z~.
\label{eq:N3_mapping_z}
\end{equation}
The corresponding transformation of the Pauli matrices $S_{\alpha,i}^x$ is given by
\begin{equation}
S_{1,i}^x = \sigma_{i}^x\,\eta_{i}^x , \quad S_{2,i}^x = \tau_{i}^x\,\eta_{i}^x, \quad
S_{3,i}^x = \sigma_{i}^x\,\tau_{i}^x\,\eta_{i}^x ~.
\label{N3_mapping_x}
\end{equation}
Inserting these transformations into the Hamiltonian (\ref{eq:HAT}) yields
\begin{eqnarray}
\label{eq:HAT_N3}
H =&& -\sum_i g_i \left( \sigma_{i}^x + \tau_{i}^x + \sigma_{i}^x\tau_{i}^x  \right)  \\
   && -\sum_i K_i \left(  \sigma_{i}^z\sigma_{i+1}^z + \tau_{i}^z\tau_{i+1}^z
                + \sigma_{i}^z\sigma_{i+1}^z\tau_{i}^z\tau_{i+1}^z  \right)    \nonumber\\
   && -\sum_i h_i \left( \sigma_{i}^x + \tau_{i}^x + \sigma_{i}^x\tau_{i}^x  \right)  \eta_i^x  \nonumber\\
   &&-\sum_i J_i \left( \sigma_{i}^z\sigma_{i+1}^z + \tau_{i}^z\tau_{i+1}^z
                + \sigma_{i}^z\sigma_{i+1}^z\tau_{i}^z\tau_{i+1}^z\right) \eta_i^z\eta_{i+1}^z \nonumber.
\end{eqnarray}
We see that the triple product $\eta_i$ does not show up in the terms containing $g_i$ and $K_i$. In the strong-coupling limit,
$\epsilon \gg 1$,
$g_i$ and $K_i$ are much larger than $h_i$ and $J_i$. The behavior of the pair variables $\sigma_i$ and $\tau_i$
is thus governed by the first two lines of (\ref{eq:HAT_N3}) only and becomes independent of the triple products $\eta_i$.
The  $\eta_i$ themselves are slaved to the behavior of the $\sigma_i$ and $\tau_i$ via the large brackets in the
third and fourth line of  (\ref{eq:HAT_N3}).

The qualitative features of the strong-coupling regime follow
directly from these observations. The first two lines of (\ref{eq:HAT_N3}) form a
two-color random quantum Ashkin-Teller model for the variables $\sigma_i$ and $\tau_i$.
As all terms in the brackets have the same prefactor, this two-color Ashkin-Teller model is
right at its multicritical coupling strength $\epsilon_c$ (as demonstrated in Ref.\ \cite{HHNV14} and shown in Fig.\
\ref{fig:2AT_PD}). The $\sigma_i$ and $\tau_i$ thus undergo a direct phase transition
between a paramagnetic phase for  $g_{\rm typ} > K_{\rm typ}$ and a ferromagnetic phase
for  $g_{\rm typ} < K_{\rm typ}$. In agreement with the quantum Aizenman-Wehr theorem, the transition
is continuous; it is in the infinite-randomness universality class of the
random transverse-field Ising model. Moreover, in contrast to the $N=2$ case, there is no
additional partially ordered phase.

What about the triple product variables $\eta_i$? For large disorder, the brackets
in the third and fourth line of  (\ref{eq:HAT_N3}) can be treated as classical variables.
If the $\sigma_i$ and $\tau_i$ order ferromagnetically, $\sigma_{i}^x + \tau_{i}^x + \sigma_{i}^x\tau_{i}^x $ vanishes
(for all sites surviving the strong-disorder renormalization group at low energies) while
in the paramagnetic phase, $\sigma_{i}^z\sigma_{i+1}^z + \tau_{i}^z\tau_{i+1}^z
+ \sigma_{i}^z\sigma_{i+1}^z\tau_{i}^z\tau_{i+1}^z$ vanishes. Thus, each $\eta_i$
becomes a classical variable that is slaved to the behavior of $\sigma_i$ and $\tau_i$.
This means, the $\eta_i$ align ferromagnetically if the $\sigma_i$ and $\tau_i$
are ferromagnetic while they form a spin-polarized paramagnet if $\sigma_i$ and $\tau_i$
are in the paramagnetic phase.

All these qualitative results are confirmed by a strong-disorder renormalization group calculation which we now
develop for the case of general odd $N$.

\subsection{Variable transformation and strong-disorder renormalization group for general odd $N$}

For general odd $N>2$, we define $N-1$ pair variables and one product of all colors
\begin{equation}
\sigma_{\alpha,i}^z = S_{\alpha,i}^z\,S_{N,i}^z \quad (\alpha=1\ldots N-1), \quad
\eta_{i}^z =  \prod_{\alpha=1}^N S_{\alpha,i}^z~.
\label{eq:Nodd_mapping_z}
\end{equation}
The corresponding transformation of the Pauli matrices $S_{\alpha,i}^x$ is given by
\begin{equation}
S_{\alpha,i}^x = \sigma_{\alpha,i}^x\,\eta_{i}^x \quad (\alpha=1\ldots N-1), \quad
S_{N,i}^x = \prod_{\alpha=1}^{N-1}\sigma_{\alpha,i}^x\,\eta_{i}^x ~.
\label{Nodd_mapping_x}
\end{equation}
In terms of these variables, the Hamiltonian (\ref{eq:HAT})  reads
\begin{eqnarray}
\label{eq:HAT_Nodd}
H=&& -\sum_i g_{i}{\left [\sum_{\alpha<\beta}^{N-1}\sigma_{\alpha,i}^x
\sigma_{\beta,i}^x+\sum_{\alpha=1}^{N-1} \prod_{k\neq\alpha}^{N-1}\sigma_{k,i}^x \right ]}
\\
 && -\sum_{i} K_{i}{\left [\sum_{\alpha<\beta}^{N-1}\sigma_{\alpha,i}^z
\sigma_{\alpha,i+1}^z \sigma_{\beta,i}^z \sigma_{\beta,i+1}^z
+\sum_{\alpha=1}^{N-1} \sigma_{\alpha,i}^z \sigma_{\alpha,i+1}^z \right ]}
\nonumber\\
&& -\sum_{i}~h_i{\left[\sum_{\alpha=1}^{N-1}\sigma_{\alpha,i}^x+\prod_{k=1}^{N-1}\sigma_{k,i}^x \right]}\eta_{i}^x
\nonumber \\
&& -\sum_{i}~J_i{\left [\sum_{\alpha=1}^{N-1}\prod_{k\neq\alpha}^{N-1}\sigma_{k,i}^z \sigma_{k,i+1}^z+\prod_{k=1}^{N-1}\sigma_{k,i}^z \sigma_{k,i+1}^z \right ]}\eta_{i}^z\eta_{i+1}^z\nonumber.
\end{eqnarray}
As in the three-color case, the $N$-product variable $\eta_i$ does not appear in the terms
containing the large energies $g_i$ and $K_i$.

We now implement a strong-disorder renormalization group for the Hamiltonian (\ref{eq:HAT_Nodd}). This can be conveniently done using
the projection method described, e.g., by Auerbach \cite{Auerbach98} and applied to the random quantum Ashkin-Teller model
in Ref.\ \cite{HHNV14}. Within this technique, the (local) Hilbert space is divided into low-energy and high-energy subspaces.
Any state $\psi$ can be decomposed as $\psi = \psi_1 +\psi_2$ with $\psi_1$ in the low-energy
subspace and $\psi_2$ in the high-energy subspace. The Schroedinger equation can then be written in matrix form
\begin{equation}
\left( \begin{matrix} H_{11} & H_{12}\\H_{21} & H_{22} \end{matrix}\right) \left ( \begin{matrix} \psi_1 \\ \psi_2 \end{matrix} \right)
 = E \left ( \begin{matrix} \psi_1 \\ \psi_2 \end{matrix} \right)
\label{eq:matrix}
\end{equation}
with $H_{ij} = P_i H P_j$. Here, $P_1$ and $P_2$ project on the low-energy and high-energy subspaces,
respectively. Eliminating $\psi_2$ from these two coupled equations gives
$H_{11} \psi_1 + H_{12} (E-H_{22})^{-1} H_{21} \psi_1 = E \psi_1$.
Thus, the effective Hamiltonian in the low-energy Hilbert space is
\begin{equation}
H_{\rm eff}= H_{11} + H_{12} (E-H_{22})^{-1} H_{21}~.
\label{eq:projected}
\end{equation}
The second term can now be expanded  in inverse powers of the large local energy scale $g_i$ or $K_i$.

In the strong-coupling regime, $\epsilon \gg 1$, the strong-disorder renormalization group is controlled
by the first two lines of (\ref{eq:HAT_Nodd}). It does not depend on the $N$-products $\eta_i$
which are slaved to the $\sigma_i$ and $\tau_i$ via the large brackets in the
third and forth lines of  (\ref{eq:HAT_Nodd}).

If the largest local energy scale is the ``Ashkin-Teller field'' $g_i$, site $i$ does not contribute to
the order parameter and is integrated out via a site decimation. The recursions resulting from (\ref{eq:projected}) take the same form as in the weak-coupling
regime, i.e., the  effective interactions and coupling strength are given by eqs.\ (\ref{eq:SDRG_Jeff}) to (\ref{eq:SDRG_eps_J})
\footnote{Strictly, (\ref{eq:SDRG_Jeff}) to (\ref{eq:SDRG_eps_J}) hold in the ground-state sector, $\tilde\zeta=1$ while
in the pseudo ground state, $\tilde\zeta =-1$, the transverse field $h_i$ shows up with the opposite sign. This difference
is irrelevant close to the fixed point because $\epsilon$ diverges (see Sec.\ \ref{subsec:observables}). Analogous statements also holds for bond decimations
and for even $N$.}.

What about the product variable $\eta_i$? The bracket in the third line of the Hamiltonian
(\ref{eq:HAT_Nodd}) takes the value $N$ while the bracket in the fourth line vanishes.
However, because $h_i \ll g_i$, the value of $\eta_i^x$ is \emph{not} fixed by the renormalization group step.
Thus $\tilde\zeta_i\equiv\eta_i^x$ becomes a classical Ising degree of freedom with energy $-N h_i \tilde\zeta_i$ that
is independent of the terms in the renormalized Hamiltonian. This means, it is
``left behind'' in the renormalization group step. Consequently, $\eta_i^x$ plays the role of the additional ``internal degree of freedom''
 first identified in Ref.\ \cite{HrahshehHoyosVojta12}.

The bond decimation step performed if the largest local energy is the four-spin interaction $K_i$ can be derived
analogously. The recursion relations are again identical to the weak-coupling
regime, i.e., the resulting effective field and coupling are given by eqs.\ (\ref{eq:SDRG_heff}) to (\ref{eq:SDRG_eps_h}).
In this step, the bracket in the fourth line of the Hamiltonian
(\ref{eq:HAT_Nodd}) takes the value $N$ while the bracket in the third line vanishes.
Thus, the renormalization group step leaves behind the classical Ising degree of freedom $\tilde\zeta_i\equiv\eta_i^z\eta_{i+1}^z$
with energy $-N J_i \tilde\zeta_i$. In the bond decimation step, the additional ``internal degree of freedom''
of Ref.\ \cite{HrahshehHoyosVojta12} is thus given by $\tilde\zeta_i \equiv\eta_i^z\eta_{i+1}^z$.

All of these renormalization group recursions agree with those of Ref.\ \cite{HrahshehHoyosVojta12} where the renormalization group was
implemented in the original variables for $N>4$ colors.

\subsection{Variable transformation and strong-disorder renormalization group for general even $N$}

For general even $N\ge 4$, the variable transformation is slightly more complicated than
in the odd $N$ case. We define $N-2$  pair variables, a product of $N-1$ colors
and a product of all $N$ colors,
\begin{eqnarray}
\sigma_{\alpha,i}^z &=& S_{\alpha,i}^z\,S_{N-1,i}^z \quad (\alpha=1\ldots N-2), \nonumber \\
\eta_{i}^z~ &=&  \prod_{\alpha=1}^{N-1} S_{\alpha,i}^z~, \quad \tau_{i}^z =  \prod_{\alpha=1}^{N} S_{\alpha,i}^z.
\label{eq:Neven_mapping_z}
\end{eqnarray}
The Pauli matrices $S_{\alpha,i}^x$ then transform via
\begin{eqnarray}
S_{\alpha,i}^x = \sigma_{\alpha,i}^x\,\eta_{i}^x\,\tau_i^x \quad (\alpha=1\ldots N-2), \nonumber \\
S_{N-1,i}^x = \prod_{\alpha=1}^{N-2}\sigma_{\alpha,i}^x\,\eta_{i}^x \tau_i^x~, \quad S_{N,i}^x =\tau_i^x~.
\label{Neven_mapping_x}
\end{eqnarray}
After applying these transformations to the Hamiltonian (\ref{eq:HAT}), we obtain
\begin{eqnarray}
\label{eq:HAT_Neven}
 H &=& -\sum_{i=1}^\infty g_{i}\left [\sum_{\alpha<\beta}^{N-2}\sigma_{\alpha,i}^x
\sigma_{\beta,i}^x+\sum_{\alpha=1}^{N-2} \prod_{\beta\neq\alpha}^{N-2}\sigma_{\beta,i}^x  + \right .
\\
 && \hspace*{3.5cm}+  \left.  \left(\sum_{\alpha=1}^{N-2}\sigma_{\alpha,i}^x+\prod_{\beta=1}^{N-2}\sigma_{\beta,i}^x \right )\eta_{i}^x \right]\nonumber
\\
&&-\sum_{i=1}^\infty K_{i}\left [\sum_{\alpha<\beta}^{N-2}\sigma_{\alpha,i}^z
\sigma_{\alpha,i+1}^z \sigma_{\beta,i}^z \sigma_{\beta,i+1}^z
+\sum_{\alpha=1}^{N-2} \sigma_{\alpha,i}^z \sigma_{\alpha,i+1}^z+ \right. \nonumber
\\
&& \hspace*{0.5cm}+ \left. \left (\sum_{\alpha=1}^{N-2}\prod_{\beta\neq\alpha}^{N-2}\sigma_{\beta,i}^z \sigma_{\beta,i+1}^z+\prod_{\beta=1}^{N-2}\sigma_{\beta,i}^z \sigma_{\beta,i+1}^z\right ) \tau_{i }^z\tau_{i+1}^z\right ]\nonumber
\\
 &&-\sum_{i=1}^\infty~h_i \left[\left(\sum_{\alpha=1}^{N-2}\sigma_{\alpha,i}^x\eta_{i}^x+\prod_{\beta=1}^{N-2}\sigma_{\beta,i}^x\eta_{i}^x \right )+ 1\right] \tau_{i}^x \nonumber
\\
&&-\sum_{i=1}^\infty~J_i\left [\sum_{\alpha=1}^{N-2}\prod_{\beta\neq\alpha}^{N-2}\sigma_{\beta,i}^z \sigma_{\beta,i+1}^z+\prod_{\beta=1}^{N-2}\sigma_{\beta,i}^z \sigma_{\beta,i+1}^z+ \right . \nonumber
\\
&& \hspace*{5.5cm} + \tau_{i }^z\tau_{i+1}^z{\Bigg]}\eta_{i}^z\eta_{i+1}^z\nonumber.
\end{eqnarray}
In contrast to the odd $N$ case, the decoupling between the pair variables $\sigma_{\alpha,i}$ and the $(N-1)$ and $N$-products
$\eta_i$ and $\tau_i$ is not complete. Each of the products is contained in one but not both of the terms that dominate
for strong coupling $\epsilon \gg 1$ (first two lines of (\ref{eq:HAT_Neven})).
As a result, the phase diagram in the strong-coupling regime is controlled by a competition between the $\sigma_{\alpha,i}^z$
and $\sigma_{\alpha,i}^x$ via the first two lines of (\ref{eq:HAT_Neven}) while the $\eta_i$ and $\tau_i$ variables are slaved to them. It features a
direct transition between the ferromagnetic and paramagnetic phases at $g_{\rm typ} = K_{\rm typ}$, in agreement with the
self-duality of the original Hamiltonian.

To substantiate these qualitative arguments, we have implemented the strong-disorder renormalization group for the Hamiltonian
(\ref{eq:HAT_Neven}), using the projection method as in the last subsection. In the case of a site decimation, i.e., if the
largest local energy is the ``Ashkin-Teller field'' $g_i$, we again obtain the recursion relations (\ref{eq:SDRG_Jeff}) to (\ref{eq:SDRG_eps_J}).
The variable $\tau_i^x$ is not fixed by the renormalization group. Thus $\tilde\zeta_i\equiv\tau_i^x$ represents the extra classical Ising degree
of freedom that is left behind in the renormalization group step. Its energy is $-N h_i \tilde\zeta_i$. If the largest local energy is
the four-spin interaction $K_i$, we perform a bond decimation. The resulting recursions relations agree with the weak-coupling
recursions  (\ref{eq:SDRG_heff}) to (\ref{eq:SDRG_eps_h}). In this case, the product $\eta_i^z \eta_{i+1}^z$ is not fixed by the
decimation step. Therefore, the left-behind Ising degree of freedom in this decimation step is $\tilde\zeta_i\equiv\eta_i^z \eta_{i+1}^z$ with
energy $-N J_i \tilde\zeta_i$.

The above strong-disorder renormalization group works for all even color numbers $N>4$. For $N=4$, an extra complication arises
because the left-behind internal degrees of freedom $\tilde\zeta_i$ do not decouple from the rest of the Hamiltonian. For example, when decimating
site $i$ (because $g_i$ is the largest local energy), the $\tau^z$ term in the fourth line of (\ref{eq:HAT_Neven})
mixes the two states of the left-behind $\tau_i^x$ degree of freedom in second order perturbation theory. An analogous problem
arises in a bond decimation step. Thus, for $N=4$ colors, the internal $\tilde\zeta_i$ degrees of freedom need to be kept, and the
renormalization group breaks down. In contrast, for $N>4$, the coupling between the internal $\tilde\zeta_i$ degrees of freedom
and the rest of the Hamiltonian only appears in higher order of perturbation theory and is thus renormalization-group irrelevant.

\subsection{Renormalization group flow, phase diagram, and observables}
\label{subsec:observables}

For color numbers $N=3$ and all $N>4$, the strong-disorder renormalization group implementations of the last two subsections
all lead to the recursion relations (\ref{eq:SDRG_Jeff}) to (\ref{eq:SDRG_eps_h}).
The behavior of these recursions has been studied in detail in Ref.\ \cite{HrahshehHoyosVojta12}.
In the following, we therefore summarize the resulting renormalization group flow, phase diagram,
and key observables.

\begin{figure}[t]
\includegraphics[width=8.5cm]{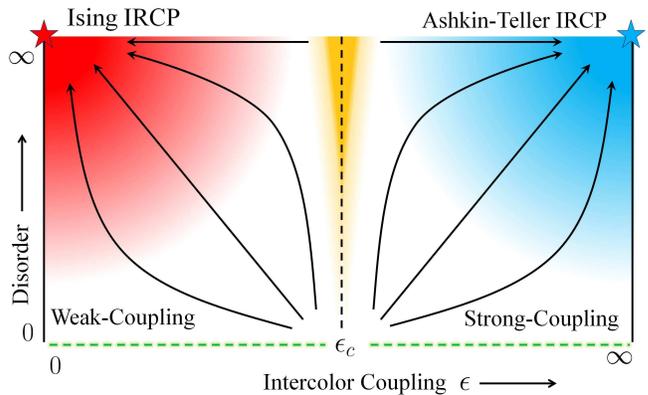}
\caption{Schematic of the renormalization-group flow diagram
on the self-duality line of the random quantum Ashkin-Teller model with $N=3$ or $N>4$ colors
in the disorder--coupling strength parameter space.
For $\epsilon<\epsilon_c(N)$ (left arrows), the critical flow approaches the
usual Ising infinite-randomness critical point of Ref.~\onlinecite{Fisher95}.
For $\epsilon>\epsilon_c$ (right arrows), we find a distinct infinite-randomness
 critical point with even stronger thermodynamic singularities (after Ref.\ \cite{HrahshehHoyosVojta12}).
}
\label{fig:rgflow}
\end{figure}

According to (\ref{eq:SDRG_eps_J}) and (\ref{eq:SDRG_eps_h}), the coupling strengths $\epsilon$ flow to
infinity if their initial value $\epsilon_I > \epsilon_c(N)$. Moreover, the competition between interactions
$K_i$ and ``fields'' $g_i$ is governed by the recursion relations (\ref{eq:SDRG_Keff}) and (\ref{eq:SDRG_geff})
which simplify to
\begin{equation}
\tilde K = \frac {K_{i-1} K_{i}}{2 (N-2) g_i}~, \quad \tilde g = \frac {g_{i} g_{i+1}}{2 (N-2) K_i}
\label{eq:SDRG_strong_coupling}
\end{equation}
in the large-$\epsilon$ limit. They take the same form as Fisher's recursions of the random transverse-field Ising model
\cite{Fisher95}. (The extra constant prefactor $2(N-2)$ is renormalization-group irrelevant).
The renormalization group therefore leads to a direct continuous phase transition between the ferromagnetic and
spin-polarized paramagnetic phases on the self-duality line $g_{\rm typ} = K_{\rm typ}$ (or, equivalently, $h_{\rm typ} = J_{\rm typ}$) . The renormalization
group flow on this line is sketched in Fig.\ \ref{fig:rgflow}.
In the weak-coupling regime, $\epsilon_I < \epsilon_c(N)$, the flow is towards the random-transverse
field Ising quantum critical point located at infinite disorder and $\epsilon=0$, as explained in Sec.\ \ref{sec:weak_coupling}.
In the strong-coupling regime, $\epsilon_I > \epsilon_c(N)$, the $N$-color random quantum Ashkin-Teller
model ($N=3$ and $N>4$) features a distinct infinite-randomness critical fixed point at
infinite disorder and infinite coupling strength. It is accompanied by two lines of fixed points for
$r=\ln(g_{\rm typ}/K_{\rm typ})>0$ ($r<0$) that represent the paramagnetic (ferromagnetic) quantum Griffiths phases.

The behavior of thermodynamic observables in the strong-coupling regime at criticality and in the Griffiths phases can be
worked out by incorporating the left-behind internal degrees of freedom $\zeta$ in the renormalization-group
calculation. This divides the renormalization group flow into two stages and leads to two distinct
contributions to the observables \cite{HrahshehHoyosVojta12}. For example, the temperature
dependence of the entropy at criticality takes the form
\begin{equation}
 S=C_1\left[\ln{\left(\frac{\Omega_I}{T}\right)}\right]^{-\frac{1}{\psi\phi}}\ln2
+C_2\left[\ln{\left(\frac{\Omega_I}{T}\right)}\right]^{-\frac{1}{\psi}}N\ln2,
\end{equation}
where $\psi=1/2$ is the tunneling exponent, $\phi = \frac 12 (1+\sqrt{5})$,
$C_1$ and $C_2$ are nonuniversal constants, and $\Omega_I$ is the bare energy cutoff.
The second term is the usual contribution of clusters surviving
under the strong-disorder renormalization group to energy scale $\Omega=T$. The first
term represents all internal degrees of freedom $\zeta$ left behind until the
renormalization group reaches this scale.
As $\phi>1$, the low-$T$ entropy becomes dominated by the extra degrees of
freedom $S\to S_{\rm extra} \sim [\ln(\Omega_I/T)]^{-1/(\phi\psi)}$.
Analogously, in the Griffiths phases, the contribution of the internal degrees
of freedom gives
\begin{equation}
\label{Sextraordered}
S_{\rm extra} \sim |r|^\nu (T/\Omega_I)^{1/(z+Az^\phi)}\ln2,
\end{equation}
 which dominates over
the regular chain contribution proportional to $ T^{1/z}N\ln2$.
Here, $\nu=2$ is the correlation length critical exponent,  and
$z=1/(2|r|)$ is the non-universal Griffiths dynamical exponent.
Other observables can be calculated along the same lines \cite{HrahshehHoyosVojta12}.

The weak and strong coupling regimes are separated by a multicritical point located at
$r=0$ and $\epsilon_I=\epsilon_c(N)$. At this point, the renormalization group flow
has two unstable directions, $r=\ln(g_{\rm typ}/K_{\rm typ})$ and $\epsilon_I-\epsilon_c(N)$.
The flow in $r$ direction can be understood by inserting $\epsilon_c(N)$ into the recursion relations
(\ref{eq:SDRG_Jeff}) and (\ref{eq:SDRG_heff}) yielding
\begin{equation}
\tilde J = \frac {J_{i-1} J_{i}}{(1 +(N-1)\epsilon_c )h_i}~, \quad \tilde h = \frac {h_{i} h_{i+1}}{(1+(N-1) \epsilon_c)J_i}~.
\label{eq:SDRG_MCP}
\end{equation}
These recursions are again of Fisher's random transverse-field Ising type (as the prefactor
$(1+(N-1) \epsilon_c)$ is renormalization-roup irrelevant). Thus, the renormalization group flow
at the multicritical point agrees with that of the weak-coupling regime. Note, however, that the
$N$ transverse-field Ising chains making up the Ashkin-Teller model do not decouple at the multicritical point.
Thus, the fixed-point Hamiltonians of the weak-coupling fixed point and the multicritical point
do not agree.

The flow in the $\epsilon$ direction can be worked out by expanding the recursions (\ref{eq:SDRG_eps_J})
and (\ref{eq:SDRG_eps_h}) about the fixed point value $\epsilon_c(N)$ by introducing
$\delta_{J,i}=\epsilon_{J,i}-\epsilon_c$ and $\delta_{h,i}=\epsilon_{h,i}-\epsilon_c$.
This leads to the recursions
\begin{equation}
\tilde \delta_J = \delta_{J,i} + \delta_{J,i+1} + Y \delta_{h_i}~, \quad \tilde \delta_h = \delta_{h,i} + \delta_{h,i+1} + Y\delta_{J,i}
\label{eq:SDRG_delta}
\end{equation}
with $Y=\epsilon_c /[(1+(N-1)\epsilon_c)(1+(N-2)\epsilon_c)]$.
Recursions of this type have been studied in detail by Fisher in the context of antiferromagnetic
Heisenberg chains \cite{Fisher94} and the
random transverse-field Ising chain \cite{Fisher95}. Using these results, we therefore find that
$\delta$ scales as
\begin{equation}
\delta_{\rm typ}(\Gamma) \approx \Gamma^{\phi_Y} \,\delta_I , \qquad \phi_Y = \frac 12 (1+\sqrt{5+4Y})
\label{eq:flow_delta}
\end{equation}
with the renormalization group energy scale $\Gamma=\ln(\Omega_I/\Omega)$. The crossover
from the multicritical scaling to either the weak-coupling or the strong-coupling fixed point
occurs when $|\delta_{\rm typ}|$ reaches a constant $\delta_x$ of order unity. It thus occurs at an energy scale
$\Gamma_x=|\delta_x/\delta_I|^{1/\phi_Y}$.

\section{Conclusions}
\label{sec:conclusions}

To summarize, we have investigated the ground state phase diagram and quantum phase transitions of
the $N$-color random quantum Ashkin-Teller chain which is one of the prototypical models for the study
of various strong-disorder effects at quantum phase transitions. After reviewing existing
strong-disorder renormalization group approaches, we have introduced a general variable transformation
that allows us to treat the strong-coupling regime for $N > 2$ in a
unified fashion.

For all color numbers $N>2$, we find a direct transition between the ferromagnetic and paramagnetic phases
for all (bare) coupling strengths $\epsilon_I \ge 0$. Thus, an equivalent of the partially ordered
product phase in the two-color model does not exist for three or more colors.
In agreement with the quantum version of the Aizenman-Wehr theorem \cite{GreenblattAizenmanLebowitz09},
this transition is continuous even if the corresponding transition in the clean problem is of first order.
Moreover, the transition is of infinite-randomness type, as predicted by the classification
of rare regions effects put forward in Refs.\ \cite{VojtaSchmalian05,Vojta06} and recently refined in
Refs.\ \cite{VojtaHoyos14,VojtaIgoHoyos14}.
Its critical behavior depends on the coupling strength.  In the weak-coupling regime
$\epsilon<\epsilon_c(N)$, the critical point is in the random transverse-field Ising universality class
because the $N$ Ising chains that make up the Ashkin-Teller model decouple in the low-energy limit.
In the strong-coupling regime, $\epsilon>\epsilon_c(N)$, we find a distinct infinite-randomness
critical point that features even stronger thermodynamic singularities stemming from the ``left-behind''
internal degrees of freedom.

The novel variable transformation also allowed us to study the multicritical point separating the weak-coupling and strong-coupling regimes.
Its renormalization-group flow has two unstable directions. The flow for  $r=\ln(g_{\rm typ}/K_{\rm typ})\ne 0$
and $\epsilon_I-\epsilon_c(N)=0$ is identical to the flow in the weak-coupling regime implying
identical critical exponents. The flow at $r=0$ in the $\epsilon$ direction is controlled
by different recursions for $\delta=\epsilon-\epsilon_c(N)$ which we have solved for general $N$.

So far, we have focused on systems whose (bare) coupling strengths are uniform
$\epsilon_{J,i}=\epsilon_{h,i}=\epsilon_I$. What about random coupling strengths?
If \emph{all} $\epsilon_{J,i}$ and $\epsilon_{h,i}$ are smaller than the multicritical value $\epsilon_c(N)$,
the renormalized $\tilde\epsilon$ decrease under the renormalization group
just as in the case of uniform bare $\epsilon$.
If, on the other hand, \emph{all} $\epsilon_{J,i}$ and $\epsilon_{h,i}$ are above $\epsilon_c(N)$,
the renormalized values $\tilde\epsilon$ increase under
renormalization as in the case of uniform bare $\epsilon$.
Therefore, our qualitative results do not change; in particular,
the bulk phases are stable against weak randomness in $\epsilon$. The same holds for
the transitions between the ferromagnetic and paramagnetic phases sufficiently far
away from the multicritical point. Note that this also explains why the randomness in $\epsilon$
produced in the course of the strong-disorder renormalization group is irrelevant if the initial
(bare) $\epsilon$ are uniform: All renormalized $\epsilon$ values are on the same side
of the multicritical point and thus flow either to zero or to infinity.

In contrast, the uniform-$\epsilon$ multicritical point itself is unstable against randomness in $\epsilon$.
The properties of the resulting random-$\epsilon$ multicritical point can be studied numerically in analogy to the
two-color case \cite{HHNV14}. This remains a task for the future.

\section*{Acknowledgements}

We are grateful for the support from NSF under Grant Nos.\ DMR-1205803
and PHYS-1066293, from Simons Foundation, from FAPESP under Grant No.\ 2013/09850-7, and from CNPq under Grant
Nos.\ 590093/2011-8 and 305261/2012-6. J.H. and T.V. acknowledge the hospitality of the
Aspen Center for Physics.

\bibliographystyle{apsrev4-1}
\bibliography{../00Bibtex/rareregions}

\end{document}